# A Constraint and Object Oriented Fifth Generation Programming Language and its Compiler and Runtime System

HAN Ji-Peng[1)]　LICHEN Zhi-Hang[1)]

[1)](Department of Technology, Beijing Huagui Technology Co., Ltd, Beijing 100081)

**Abstract**　Since the advent of LISP, the fifth generation programming language has developed for decades. However, compared with the fourth generation programming language, the fifth generation programming language has not been widely used because of its obscure semantics, rigorous representation of problems, and limited inference ability. For this reason, COOL (Constraint and Object Ordered Language), a fifth generation programming language proposed in this paper, overcomes the problems of intuitive semantics, rigorous restrictions on handling problem conditions, and improves the inference ability of the fifth generation programming language. Specific improvements are as follows: First, COOL supports process-oriented and object-oriented for easy application in production projects; Second, COOL supports expression as function declaration and function return, which improves language affinity for mathematical formulas, and supports embedding function parameters into function name strings to make function naming closer to natural languages; Third, COOL introduces a weighting mechanism and accelerates the inference process through cumulative weighting. Fourth, COOL introduces the concepts of forward and reverse functions in programming so that computers can infer and execute problems with logical sequential constraints. Fifth, the computer can deduce the reverse solution process by using the forward solution process through the back-tracking algorithm and the dynamic programming algorithm, so that the computer can deduce the problem with time-sequential constraints. Sixth, the pre-execution step is introduced to separate the inference and function query process of the program from the execution process, so as to improve the execution speed of the program.

**Key words**　The fifth generation programming language; Object-oriented; Automatic reasoning; Reverse reasoning; Cumulative weight; Dynamic planning; Pre-Execution;

## 1 INTRODUCTION

The development of programming language has experienced machine language, assembly language, advanced language, function language, logical language[1]. As the fifth generation programming language, logical language can theoretically reduce the burden of user thinking through computer reasoning, so as to improve the efficiency of programming. However, in actual production, no logical programming language has become the mainstream production tool so far due to its limitations in scope and performance.

The oldest logical language to date is the list processing language LISP (LISt Processing), created by John McCarthy in 1958. It is characterized by a symbolic expression rather than a number, where the symbolic expression is represented by a list structure or an S expression. Compared with other earlier programming languages, LISP can still be used as a convenient tool for developing higher-level symbolic computing systems and artificial intelligence because of its runtime system with easy access to host and operating system functions, and its internal language as a list structure for high-level language compilation purposes. But at the same time, LISP's weakness in numerical operation and unclear mathematical semantics restrict its further widespread use[2].

In 1975, the PROLOG (PROgramming in LOGic) developed by the Kowalskl, Colmerauer and van Emden teams at the University of Marseilles was launched. PROLOG is a programming language based on symbolic logic. Its basic computer mechanism is a pattern matching process. Pure LISP can be thought of as a special tool for PROLOG whose processes are limited to simple functions and data structures are limited to lists. For users, the language is clear, readable, concise, and easier to program[3]. However, there is a point of view that PROLOG is actually a deductive inference technology based on reverse rules, but with strict restrictions on the representation of rules and objectives. Combined with the simplicity of deductive inference control mechanism itself, it is difficult to apply to complex application environment[4].

From 1982 to 1993, Japan carried out research work on the Fifth Generation Computer Systems (FGCS), the technical goal of which was to develop parallel inference technology. Researchers have attempted to develop a kernel language that can express and execute logical programming of various parallel algorithms[5]. However, the project has generally failed because its objectives are too ambitious and some initial features such as machine translation, pattern recognition have not been achieved[6]. The successful application of neural networks in machine translation and pattern recognition further compresses the application space of logical programming in artificial intelligence development.

In 1987, the Jaffar team proposed the Constraint Logic Programming(CLP) language, which refers to the basic components of a problem as constraints and uses rules to combine constraints to represent the whole problem. The CLP language has strong expressive power because the user directly manipulates various constraints. Early representations of the CLP language were CAL[8] and CHIP[9].

Currently, the main research directions of logical programming languages are Probability Logic Programming Language PROBLOG (PROBabilistic LOGic) and Constraint Handling Rules (CHR), both of which introduce probability into logical programming.

Probability Logic Programming Language mainly studies the probability of inferring and calculating results from given rules by labeling certain facts with probability. Constraint Processing Rule Language CHR was developed from the Constraint Logic Programming Language CLP and was originally designed by Fr hwirth in 1991 to write a user-defined constraint solver[10]. Later, the CHR language derived the Probability Constraint Handling Rules language (PCHR), which allows users to "probability weight" rules to determine the probability that they will be applied in inference[11].

The development of Constraint Logic Programming and Probability Logic Programming has increased the reasoning capabilities of the fifth generation programming languages, and has widened the range of problems they can handle. However, their difficulties in practical production have not yet substantially improved. And Philip Wadler et al. pointed out some problems in logical programming and functional programming:

The languages are less efficient and do not have good debuggers.

A large number of constraints are required in complex relationships.

Unfamiliarity with the languages makes it difficult to sell and maintain products.

Lack of fully functional libraries and interfaces with other languages;

Researchers in programming languages focus on system development rather than application development.

Most importantly, functional programming languages and logical programming languages are preferred only within a narrow range of theorem proving, lacking killer applications that can attract a large number of users[12].

In addition, the author believes that the application of logical programming language is limited for the following reasons:

A large number of symbolic representations may be concise for researchers, but may not be intuitive for most users;

The development of neural network has weakened the advantage of logical programming in computer reasoning. Current logical programming languages are difficult to develop neural network.

The narrow scope of application makes the logical programming language can only be used to develop individual modules in the software, and the huge difference between the logical programming language and the common high-level language grammar style causes the overall development cost of the project to increase.

This paper presents an easy-to-use fifth generation programming language COOL by analyzing, referencing and improving the characteristics, advantages and disadvantages of some current representative logical programming languages. This language has the main features of CHR, and it also absorbs the design ideas of PCHR and PROBLOG which have been studied extensively in recent years, that is, introducing probability into "rules" and "facts" to calculate and control. Probability will be replaced with weight in this language. In order to be closer to the actual production process, this language abandons the LISP list style and PROLOG-like language style of the traditional logic language, and changes to the C++/JAVA style to adapt to the mainstream software engineering development process, reduces the learning cost of users and enables the modular development of rule libraries. For a piece of code, the order in which the statements are executed also constitutes a constraint. This language solves this constraint by dynamic programming, and implements reverse inference for some functions. To better describe the semantics of mathematical expressions, this language supports expression as function declaration. In fact, all function declarations in this language are stored in expression form (tree structure). Because the inference mechanism and function query mechanism of this language need to match the tree structure and involve a lot of traversal operations, we separate this part of the operation from the execution process, which we call "pre-execution", "pre-execution" process is similar to Haskell's lazy operation[13], but has a special set of execution rules. Users can combine pre-execution steps with compilation steps to reduce code execution time or with execution procedures to reduce code volume based on specific circumstances in practice. The main contributions of this paper are as follows:

• The design idea and specific framework of a Constraint and Object Oriented programming language are presented.

• A weight-based inference control method is presented for the constraints inference process in a Constraint and Object Oriented programming language.

• Reverse reasoning for simple sequential structure functions is implemented through dynamic programming.

• Presents a pre-execution step that separates the inference process and function query process from the execution process.

## 2 RELATED WORK

COOL is a fifth generation language with CHR language characteristics. From the user writing the program to the final result, the overall process is: coding (functions, variables, control structures, related syntax of classes), precompilation, compilation, pre-execution, execution. This section will be introduced step by step according to the overall process sequence.

### 2.1 FUNCTION

Functions are the most important part of COOL. This section first provides some simple code to familiarize readers with the style of COOL functions, then introduces functions that return expressions and return operands, functions with additional weights, forward and reverse functions, function weights, function inverses. These constitute the COOL inference framework.

#### 2.1.1 DEFINE FUNCTION

In COOL, a function consists of a function declaration and a body of functions. A function representing addition can be defined as code 1:

**CODE 1** Example function declaration
$@add(a,b)\{$
    $return: a + b;$
$\}$

Where '@' modifies the subsequent $add(a,b)$ as a function declaration rather than a function call. You need to remove '@' when calling a function:

**CODE 2** Function call example
$add(1,2);$

Functions are called with preference for passing in references to actual parameters rather than copying, for example, code 3:

**CODE 3** Example function call
$@add(a)to(b)\{$
    $b = b + a;$
$\}$
$add(1)to(x);$

Increase $x$ by 1 after code 3 is executed. In this code, function parameters are allowed to be embedded in function name (predicate) string to enhance expressiveness.

In addition, COOL provides built-in functions that can be used without definition to provide basic mathematical operations.

#### 2.1.2 FUNTION RETURNING EXPRESSION and FUNCTION RETURNING VALUE

The attribute "exp" is added to the function declaration to indicate that the return of this function is an expression. This kind of function is used to express the transformation rules of the expression. Users cannot directly call the function returning expression.

For example, code 4 describes the inverse operation of the distributive law of multiplication by function:

**CODE 4** Example of function returning expression
$exp: @\{a * c + b * c\}\{$
    $return: (a + b) * c;$
$\}$

The pair of curly braces immediately following "@" and the expression within them are called function declaration scope, and the only internal expression is called function declaration expression. In fact, in the function example in section 2.1.1, the function name is also in its function declaration scope, but the curly brackets on the scope boundary are omitted for writing convenience.

If the function declaration has no attribute "exp", the function returns the computed value. The user can only call value returning function directly.

### 2.1.3 FORWARD FUNCTION and REVERSE (BACKWARD) FUNCTION

Both forward and reverse functions are for value returning functions; Functions returning expression do not have this property.

Forward function refers to the function whose parameters are determined and whose return value is undetermined. The execution process of forward function is to deduce the return value of the function by using the input parameters, such as code 1.

Reverse function refers to the function whose return value is determined and whose input parameters have pending parameters (parameters with undetermined values). The execution process of reverse function is to use the return value of the function and the determined input parameters to calculate the pending input parameters.

For example, code 5 provides the reverse function required to calculate a solution of a quadratic function:

CODE 5 Example reverse function
$@\{a*\$x^2 + b*x + c\}\{$
    $x = (-b + (b^2 - 4*a*(c-ans))^0.5)/(2*a);$
$\}$
$@\{\$a == b\}\{$
    $a = b;$
$\}$

The "$" symbol acts on parameter x, indicating that parameter x is pending in the expression. For variables that appear multiple times in an expression, you only need to use "$" to decorate them once. The return value of the function represented by the variable "ans" appearing in the function body is a known parameter, which you can use or not use.

When calling the reverse function, the user needs to modify the undetermined variable to be derived with "$", for example, solve the univariate quadratic equation with $a$ as the unknown number (code 6):

CODE 6 Example for calling reverse function
$1*\$a^2 + (-2)*a + 1 == 0;$

$a$ needs to be decorated with "$".

### 2.1.4 FUNCTION WEIGHT

The function weight is used to control the reasoning direction of the reasoning system. Users can give greater weight to transformations (i.e. functions) that are easier to lead to correct reasoning results, and less weight to transformations that are not easy to lead to correct reasoning results. The weight of a function is determined at the time of function declaration, between "@" and the scope of function declaration. For example, code 7 defines a function with a weight of 10:

CODE 7 Examples of function with weight
$@(10)\{\$a == b;\}\{$
    $a = b;$
$\}$

A function that is not weighted (such as code 1) has a weight of 0.

### 2.1.5 FUNCTION INVERSION

Function backstepping is a special way to define reverse function through forward function. Example code 8:

CODE 8 Example of using forward function to define reverse function
$@\ price\ of\ buying\ (a)\ kg\ of\ apple\ unit\ price\ (b)\{$
    $return: a*b;$
$\} => @apple\ unit\ price\ (b)\ can\ be\ bought\ (\$a)\ kg;$

Where "=>" indicates derivation. The necessary conditions for using this method to define the reverse function in the algorithm to realize this function in this paper are: the names of all parameters in the function declaration of the reverse function must correspond to the names of all parameters in the function declaration of the forward function one by one; The undetermined parameter in the reverse function and the variables that depend on this parameter only participate in the sequential structure of the forward function, and do not participate in the loop and branch structure of the function body; The function body code of forward function does not modify the parameters outside the scope; Variables with the same name and different scopes do not exist in the function body of forward function. Since it is a necessary condition, it also means that even if these requirements are met, the backstepping may not be completed for other reasons.

## 2.2 VARIABLE

### 2.2.1 DECLARATION and TYPE of VARIABLE

Non temporary variables of COOL must be declared before use. An example of declaring variable $a$ and assigning an initial value of 1 is shown in code 9:

CODE 9 Example of variable declaration

```
new: a = 1;
```

The type of the variable is determined by the last assigned type. The basic data types supported by COOL are floating point numbers and strings; When a variable is not assigned a value, its type defaults to floating number.

2.2.2 ACCESS to VARIABLES

Variables can be accessed from the point where they are declared to the end of their scope. When an expression needs to use a variable, it takes precedence over the variable in the current scope. Users can use "out" to modify this variable to force the expression to use variables from the upper scope. For example, code 10:

**CODE 10** "out" usage example
```
new: a = 1;
{
    new: a = 0;
    a = out: a + 1;
}
```

The result of calculation (the final value of $a$ in the inner scope) is 2.

When the "out" modifier parameter is used in a function declaration scope, it is no longer a formal parameter but an actual parameter outside the scope of the function declaration. For example, $pi$ in the function declaration of code 11:

**CODE 11** Example use of "out" in a function declaration
```
new: pi = 3.14159;
   … …
exp: @{sin(out: pi/2 − a)}{
    return: cos(a);
}
```

## 2.3 CONDITIONAL STATEMENT

### 2.3.1   LOOP

COOL supports "while" loop structure, such as code 12:

**CODE 12** "while" loop example
```
while(a > 0){
    … …
}
```

### 2.3.2   BRANCH

Branch structure of COOL, example code 13:

**CODE 13** Branch structure example
```
if(a == 0){
    … …
}elif(a > 0){
    … …
}else{
    … …
}
```

## 2.4 COMMENT STATEMENT

The comment style is the same as C, for example, code 14:

**CODE 14** Example of comment
```
//Single − line comments
/∗ One or more lines of comments ∗/
```

## 2.5 COMPLETE CODE EXAMPLE 1

Solve the following mathematical problems:

For two numeric quantities $x$ and $y$, do the following in turn:

1. Sum $x$ and $y$ and record the result as $a$;
2. Modify the value of $x$ so that it satisfies the value of $x + 1$ equal to $y$;
3. Solve the variable $z$, where $z^2 + x * z + y$ equals 100;
4. Sum $a, x, z$ to get the final result;

Given that the result from the fourth step is 50 and that $y$ has an initial value of 3, what is the initial value of $x$?

The complete COOL code for solving the problem, such as code 15, involves function returning expression and value returning function, forward and reverse functions, function weights, and function inverse.

In the following section, it is used as an example code to show the reasoning mechanism:

**CODE 15** Complete code example 1
```
/∗ Function for solving quadratic equations ∗/
@(100){a ∗ $x^2 + b ∗ x + c}{
    x = (−b + (b^2 − 4 ∗ a ∗ (c − ans))^0.5)/(2 ∗ a);
}
/∗ Defines the law of additive exchange a + b →b + a, and
the function declaration is modified by attribute "exp",
indicating that this function is a function returning
expression. For the function returning expression, the
function parameters a and b are modified by" #", indicating
that the law of additive exchange can be applied regardless
of whether a or b is undetermined. ∗/
exp: @(−1){#a + #b}{
    return: b + a;
}
/∗ Define addition − subtraction conversion：
    a − b →a + (−b);∗/
exp: @(−1){#a − #b}{
    return: a + (−b);
}
/∗ Define additive reverse function ∗/
@(10){$a + b}{
    a = ans − b;
}
/∗ Define reverse function of equation ∗/
@(10){$a == b;}{
    a = b;
}
∗ Define a forward function for deriving the reverse
function ∗/
@get result from (x) and (y){
    new: a = x + y;
```

```
    $x + 1 == y;
    new: z = 0;
    1 * $z^2 + x * z + y == 100;
    return: a + x + z;
} ⇒ @get result from ($x) and (y);

new: x = 0;
new: y = 3;
get result from ($x) and (y) == 50;
x --> 0;/* "-->" indicates output, which means that
the x value is output by default */
```

## 2.6 CLASS

Rules for solving similar problems can be encapsulated into classes. Users can reuse, modify and expand rules more flexibly through inheritance, so as to realize the division and treatment of complex problems and the modular development of programs.

### 2.6.1 DEFINE CLASS

In COOL, a class consists of a declaration and the scope of the class (hereinafter referred to as the "class body"), such as code 16:

**CODE 16** Example of class
```
system : OperationLaw{
    … …
}
```

Where, "system" is the keyword of the declaration of class; "Operationlaw" is the name of the class.

### 2.6.2 INHERIT

When defining a class, you can make it inherit from other classes to use its member functions and variables:

**CODE 17** Class inheritance example
```
system: MainProcess <<
OperationLaw, QuadraticEquation {
    … …
};
```

Where "<<" means inheritance. When a class inherits multiple classes, the names of the inherited classes are separated by commas. If a variable with the same name or a function with the same declaration in the parent class exists in the current class, the member variable or function of the current class will be used first by default; If multiple parent classes have the same member, the member of the class on the left at the time of declaration is preferred (for example, if the class $Operationlaw$ has the same member as the class $Quadracutequation$, the member of $Operationlaw$ is preferred); If the name of the class is specified, the member function of the specified class is used.

### 2.6.3 CLASS INSTANCE INITIALIZATION

An instance of a class also belongs to a variable. The way to declare it is shown in code 18:

**CODE 18** Class instance declaration example
$MainProcess: m;$

Its initialization is similar to a function call. After entering the scope of the class, it creates an active record and executes the code in the scope of the class in turn. However, when leaving the scope of the class, it does not destroy the active record, but takes the active record as the value of the corresponding variable.

### 2.6.4 ACCESS MEMBERS

By "." Operator to access member variables or call member functions, such as code 19:

**CODE 19** Example of access members
```
m.x = 1;
m.constructor();
```

## 2.7 COMPLETE CODE EXAMPLE 2

Code 20 shows the combined use of cool's classes.

First, two classes are defined. The class $Operationlaw$ contains the transformation rules of some common operations. The class $Quadricequation$ contains a formula for solving the univariate quadratic equation. Then, the main program class $Quadricequation$ inherits the two classes previously defined, and defines two functions in the program to solve the univariate quadratic equation and modify the member variables:

**CODE 20** Complete code example 2
```
//Define classe composed of operation laws
system: OperationLaw{
    //transposition of terms
    exp: @(-10){$a == b}{
        return: a - b == 0;
    }
    //Addition and subtraction conversion
    exp: @(-10){#a - b}{
        return: a + (-b);
    }
};
//Define classes for solving quadratic equations
system: QuadraticEquation{
    /* Functions for solving standard quadratic
    equations of one variable */
    @(-100){ a * $x^2 + b * x + c == 0; }{
        x = (-b + (b^2 - 4 * a * c)^0.5)/(2 * a);
    }
}
/* "<<" means inherit. The MainProcess class
inherits the OperationLaw class and the
QuadraticEquation class. It can access the members of
the two classes at the same time */
system: MainProcess <<
OperationLaw, QuadraticEquation {
    new: x = 1;
    @constructor(){
        1 * $x^2 + 4 * x == 100;
        x --> 0;
    }
```

```
        @increase(n){
                x = x + n;
        }
};
MainProcess: m;//create class objects
m.constructor();//Call member function
m.increase(10);
m.x−→ 0;//Output the value of the member variable
```

## 2.8 PRECOMPILE

The precompile process performs the following operations:

・Delete source code comments;

・Delete spaces and line breaks in the source code;

・Merge the source code in multiple files;

・Code replacement;;

・Delete invalid source code; Wherein, the invalid source code is the code that is marked by the user through the preprocessing instruction that does not need to be compiled and the code that will not be executed due to the non-standard writing by the user;

・Replace the non ASCII code appearing in the source code identifier with ASCII code string

・Deform the function declaration and its call statements composed of strings and parameters in the source code; In this step, move all function parameters that are not after the function name string to the end of function name string, and keep the order of function parameters unchanged; At the same time, a new string is formed by replacing the original position of the function parameter with a specific string, which is used to identify a parameter originally corresponding to this position, for example: "_ARG_"; Specifically, $add\ (a)\ and\ (b)\ to\ (c)$ is transformed into $add\_ARG\_and\_ARG\_to(a,b,c)$ after precompiling. This method allows function parameters to embed function name string.

## 2.9 COMPILE

The precompiled code is compiled into character code through lexical analysis program and syntax analysis program. The character code consists of three parts, including code type flag bit, quaternion and quaternion parameter type flag bit array.

Code types and the functions they describe include:

"Variable declaration", declaring a variable;

"Function operation", including the operation of binding the scope of function declaration to the scope of function body and the operation of binding weight and return type to the scope of function declaration;

"Derivation function", deriving the implementation logic of reverse function according to the function body scope of forward function;

"Scope start" is used to declare the start of the scope;

"Scope end" is used to declare the end of the scope;

"Expression", indicating that the quaternion of this line of code is part of the expression; Operations, function calls, and assigning attributes ("$", "out") to variables are "Expressions";

"Jump", when the condition is true, jump to the code corresponding to the address for execution;

"End of expression" indicates that the expression of the code in the previous line of the current executing code has ended, which plays the role of truncating the expression;

"Return" is used for function return;

"Access member", the computer is required to access parameters according to the given path;

"Class operation", including declaring a class, binding the class to the scope, and class inheritance;

Code types and the functions they describe include:

"Variable declaration", declaring a variable;

"Function operation", including the operation of binding the scope of function declaration to the scope of function body and the operation of binding weight and return type to the scope of function declaration;

"Derivation function", deriving the implementation logic of reverse function according to the function body scope of forward function;

"Scope start" is used to declare the start of the scope;

"Scope end" is used to declare the end of the scope;

"Expression", indicating that the quaternion of this line of code is part of the expression; Operations, function calls, and assigning attributes ("$", "out") to variables are "expressions";

"Jump", when the condition is true, jump to the code corresponding to the address for execution;

"End of expression" indicates that the expression of the code in the previous line of the current executing code has ended, which plays the role of truncating the expression;

"Return" is used for function return;

When "member access", the computer is required to access parameters according to the given path;

"Class operation", including declaring a class, binding the class to the scope, and class inheritance;

Quaternion and quaternion parameter type flag bit array are used to store the left operand, right operand, operator, operation result and corresponding data type of quaternion when the code type is "Expression"; When the code is of other types, it is only used to store relevant parameters and corresponding data types.

## 2.10 PRE-EXECUTION

### 2.10.1 LOAD CHARACTER CODE

In the pre execution process, the compiled character code file will be loaded first, and it will be converted into extended code line by line. At the same time, the conversion will generate the scope table, function table and class table to make it easier to obtain the required information.

The specific contents of each table generated during the loading process are as follows:

(1) Extended code table

Extended code table is a two-dimensional table with code address as key and extended code as value; The key-value pairs (tuples) in the table are arranged from small to large and from front to back according to the address;

The expanded code specifically includes: address of the scope where the code is located, code executable flag bit, code type flag bit, quaternion parameter type flag bit array, quaternion formal parameter flag bit array, quaternion local scope parameter flag bit array, quaternion parameter pending flag bit array, bound function address, function root node flag bit. The functions of each part are:

address of the scope where the code is located：Used to identify the address in the scope table of the scope where this line of code is located;

code executable flag bit：It is used to identify whether the code can be executed. The code corresponding to the scope of the function declaration cannot be executed. This flag bit is "false"; The function body scope, class scope and the jump part of the conditional statement can be executed only after certain conditions are met. This flag bit is "execute under certain conditions"; In other cases, this flag bit is "true";

Code type flag bit, quaternion , quaternion parameter type flag bit array: the same functions as those in character codes;

Quaternion formal parameter flag bit array: valid when the executable flag bit of the code is "false" (that is, the code is in the scope of the function declaration), which is used to identify whether the parameter in the quaternion is a formal parameter;

Quaternion local scope parameter flag bit array: shares the storage space with quaternion formal parameter flag bit array. It is valid when the executable flag bit of the code is not "false" (that is, the code is not in the scope of the function declaration), It is used to identify whether to query the corresponding parameters in the quaternion from the scope of the current code (flag bit is "true") or from the upper scope of the scope of the current code (flag bit is "false");

Quaternion parameter pending flag bit array: used to identify whether the value of the corresponding parameter in the quaternion is pending when this line of code is executed. If the parameter is pending (the flag bit is "true"), the value of this parameter will be regarded as unknown. Even if this parameter has a corresponding value at this time, its value will be regarded as invalid; If the parameter is determined (the flag bit is "false"), the value of this parameter will be regarded as a known condition; In particular, for the function returning expression, in many cases, the rules it represents can be used regardless of whether the variable is pending, so the parameters in the function declaration are allowed to have three states: "pending", "determined", and "unrelated". Accordingly, the flag bits also have three states: "true", "false", and "unrelated";

Bound function address: The address used to identify the function bound to the code (segment); Since all function declarations in COOL are expressions, calling a function requires that the function declaration expression of the function be bound to the expression with the same structure, and the bound function is called when the expression corresponding to the code snippet is executed. The address was determined in step 2.10.2;

Function root node flag bit: This flag bit is true, indicating that this line of code corresponds to the root node of the function declaration expression of the bound function and false to the child node. Used to handle situations where a piece of code is bound to the same function multiple times in a row (that is, the same function is called in a row), and since each line of code has the same bound function address, it is not possible to directly differentiate between individual function calls. This flag is determined at step 2.10.2.

（2）Scope table

The scope table is a two-dimensional table with the address of the scope as the key and the scope information as the value. Each key-value pair in the scope table is arranged from small to large and from go to back according to the address.

Scope information includes scope type flag bit, scope start address, scope end address, and address of scope used for parameter query. The functions of each part are:

Scope type flag bit: used to describe a scope as a function declaration scope, or, a function body scope;

Scope start address: The address of the scope start code in the code table;

Scope end address: The address of the scope end code in the code table;

Address of scope used for parameter query: Address of the scope to be accessed when querying parameters that are involved in but not declared in this scope and in the outer scope of this scope. "Address of scope used for parameter query" of the function implementation (body) scope is the corresponding function declaration scope address;

（2）Function table

Function table is a two-dimensional table with function address as key and function information as value. The key-value pairs in the function table are arranged from small to large and from going to back according to the size of the address.

Function information includes: function declaration scope address, function declaration root node address, function body scope address, "return" code address array, function returning expression flag bit, reverse function flag bit, function weight; The functions of each part are:

Function declaration scope address：Identifies the scope address of the function declaration;

Function declaration expression root node address: address of the code where the root node quaternion of function declaration expression is located;

Function body scope address: address of function body scope

"Return" code address array: code addresses of all return statements in the scope of the function body;

Function returning expression flag bit：used to identify the function return expression or return operation value;

Reverse function flag bit: used to identify the function as forward function or reverse function

Function weight: identifies the weight value of a function;

（3）Class table

Class table is a two-dimensional table with class address as key and class information as value; The key value pairs in the class table are arranged from small to large and from front to back according to the size of the address;

The address of the class is the address of the scope of the class body;

Class information includes: class scope address, class name and parent class address array; The functions of each part are:

Class scope address: the address of the scope of the class body.

Class name: name of the class.

Parent class address array：an array that stores the addresses of classes directly inherited by the class.

During loading, all addresses are converted to multi-level addresses to facilitate subsequent code table insertion and deletion. Multi level addresses meet:

Multi level address is an address composed of a one-dimensional integer array and the first and last bits of the array are not zero; The multi-level address with only one number in the array is the single-level address, and the multi-level address with zero numbers in the array is the empty address; The size comparison method between multi-level addresses is to start from the starting position of the multi-level address array and then compare their corresponding digits backward: if each digit is equal, the two multi-level addresses are equal; if two digits are not equal, the address corresponding to the array where the larger digit is located is larger; If the array of one multi-level address is shorter than the array of another multi-level address, the number of the empty position of the shorter array relative to the longer array is zero by default.

2.10.2  REASONING PROCESS

(In this section, for convenience of expression, code segments, expressions and syntax trees are allowed to refer to each other without ambiguity.)

From front to back, execute each line of code in the extended code table according to the following rules:

Step A: for the code whose "code executable flag bit" is "false", directly skip the current execution code;

Step B: for the code whose "code executable flag bit" is "execute under certain conditions" or "true", execute one or more of the following sub steps according to the "code type flag bit":

Sub step B-1: when "code type flag bit" is "jump", skip the current execution code;

Sub step B-2: when "code type flag bit" is "return", skip the current execution code;

Sub step B-3: when "code type flag bit" is "function operation", skip the current execution code;

Sub step B-4: when "code type flag bit" is "scope start", create an active record for the scope starting from the currently executed code in the current active record, and mark the created active record as the current active record; In particular, when this scope is the scope of a class, if it has an inherited parent class, the pointer to the active record of the parent class is directly added to the "array of pointers to parameter query active records" of the current active record;

The active record is used to record the context information involved before and after the specific execution of the code within a scope and the information generated during the execution, specifically including: data table and active record information;

The data table is a two-dimensional table. The key of the two-dimensional table is the identifier name, or the address of the parameter, or the address of the scope of the active record. The value of the two-dimensional table is a number, or a string, or an active record, or a path to access another parameter, or other supported data types.

The active record information specifically includes: formal parameter and actual parameter cross reference table, active record scope address, pointer to parent active record, pointer to return to active record, return address of current active record, array of points to parameter query active records;

The "formal parameter and actual parameter cross reference table" mentioned above is a two-dimensional table with the formal parameter identifier as the key and the actual parameter identifier or the actual parameter address as the value. This table is created when and only if the active record is an active record of a function body scope; During the pre-execution process, this table is only obtained by analyzing the function declaration scope, only the formal parameters are stored, and the actual parameters corresponding to the formal parameters are empty (default value); During execution, this table is obtained by comparing the function declaration scope with the code segment that calls this function to store the formal parameters with the corresponding actual parameters.

The active record scope address is the address of the scope corresponding to this active record;

The parent activity record described refers to an active record whose data table range directly contains this active record.

The return to active record described refers to an active record that should be set as the current active record after it has left the current active record, and the return to active record of an active record defaults to its parent active record. If an active record is created as a result of a function call, its return to active record should be an active record that records the action of the function call. When querying the actual parameters, you need to visit return to active record for query.

The return address of current active record refers to the address of the code that should be marked as currently executed when returned from the current active record, defaulting to the next line of code currently executed. If the current active record was created as a function call, the return address should be the next line of code at the function call.

The parameter query active record described refers to the active record that needs to be accessed when querying variables that do not appear in the data table of the current active record and the formal parameter and actual parameter cross reference table. The parent instance generated when a subclass instance is created is the parameter query active record corresponding to the active record for this subclass instance.

Sub step B-5, when "code type flag bit" is "Scope end", destroy and leave the current active record if the current scope is not a class instance, and mark return to active record as the current active record; If the current scope is a class instance, leave the current active record without destroying it, and mark return to active record as the current active record.

Sub step B-6, when "code type flag bit" is "Variable declaration", add a key-value pair with the parameter identifier as the key and the default value of the parameter to the data table of the current active record. In particular, when a declared variable is an instance of a class, an active record is no longer created for that class's instance, but directly references the existing active record corresponding to the class scope, which means that all instances of the same class use the same active record during pre-execution.

Sub step B-7, when "code type flag bit" is Expression and the current code is not scoped within the body of a function returning expression, execute any of the following sub steps:

Sub step B-7-1 skips the current code directly when the currently executed code is bound to a function, that is, when the bound function address of the currently executed code is not empty;

Sub step B-7-2, when the code currently executed is not bound to a function, that is, its bound function address is empty, performing code replacement and function binding operations;

In this sub-step B-7-2, a data representation format $a(b|c)$ is defined to indicate that data $a$ is created or assigned in the $b$-cycle (number of matching rounds) and used in the $c$-cycle. Define the representation of a key-value pair $<a,b>$, indicating that the key is $a$ and the value is $b$.

In this sub-step B-7-2, the conditions for initialization

(Round 0 Matching), Round $k$ Matching, and Exit Matching are done in turn. Specifically, this sub-step B-7-2 performs the following operations:

B-7-2-1, copy the code segment $e$ corresponding to the expression with the longest code currently executed, and get the key-value pair $<0, e>$ with the initial cumulative weight value of 0 as the key and the copied code segment as the value.

B-7-2-2, add a duplicate code segment key-value pair to the current code table silo $S$, where $S$ is a two-dimensional table with a cumulative weight value as the key and a code segment as the value, and is also considered a set of key-value pairs;

$$S(0|1) = <0, e> \qquad (1)$$

At this time, $S(0|1)$ only has $<0, e>$ as an element;;

The code table silo described is a two-dimensional table with the cumulative weight value as the key and the code segments as the value. Each key-value pair in the code table silo is arranged from small to large according to the cumulative weight value. The maximum number of stored key-value pairs $m$ in the code table silo is set by the user.

Optionally, adding elements to the code table silo meets the following requirements a and b:

a: When attempting to add a new key-value pair to a full code table silo, if the cumulative weight value in the key-value pair is greater than the minimum value of the cumulative weight value in all keys of the code table silo, the code table silo first deletes its header key-value pair and then adds a new key-value pair, if the cumulative weight value in the key-value pair is less than or equal to the minimum value of the cumulative weight value in all keys, Code table silo will not do anything with the added key-value pairs, nor will it add new key-value pairs;

b: Code table silo allows the existence of key-value pairs with identical keys and different values, and does not allow the existence of two identical key-value pairs.

. For $k \in N^*$, when the current code table silo $S(k-1|k)$ is not empty, for all $<w_i, e_i> \in S(k-1|k)$, $\{i|i \in N \land i < card(S(k-1|k))\}$; Match all syntax tree branches (corresponding sub code segments) in $e_i$ that meet the requirement that no function is bound to the code of any node, with all accessible functions. The functions that are accessed first are matched first. Whenever a syntax tree branch of $e_i$ matches a function $fun$ successfully, code replacement or function binding is performed on $e_i$ according to the type of function to generate code segment $e_j$; The corresponding weight value of $e_j$ is the result of adding the weight value of $fun$ to the weight value of $e_i$, namely:

$$w_j = w_i + w_{fun} \qquad (2)$$

The accessible functions refer to the functions declared before the current location and located in the current scope or the upper scope of the current scope, and if the scope of the current code is the scope of a class or its child scope, the member functions of the inherited classes of this class can also be accessed; Code segment matching function refers to comparing the expression $exp1$ represented by code segment with function declaration expression $exp2$. If the syntax tree structures of $exp1$ and $exp2$ are the same, and any node $node1$ of $exp1$ and $node2$ of $exp2$ that is aligned with $node1$ are taken, the matching is successful if $node1$ and $node2$ meet one of the following five conditions:

Node1 node2

• Node1 is a concrete value and node2 is a formal parameter of the same type

• Node1 is a variable and node2 is a formal parameter of the same type

• Node2 is the actual parameter and node2 and node1 are the same variable, for example, code 21:

• $node1$ is a specific value and $node2$ is the same value

• $node1$ is a specific value and $node2$ is a formal parameter of the same type

• $node1$ is a variable and $node2$ is a formal parameter of the same type

• $node2$ is the actual parameter and $node2$ and $node1$ are the same variable, for example, code 21:

**CODE 21** Node matching example
```
new: N = 3;
@{f(N, a)}{
    return: N − a;
}
@{f(out: N, a)}{
    return: N + a;
}
f(N, 4);/* Successfully matched with  @{f(N,a)} 
and @{f(out: N, a)} */
{
    new: N;
    f(N, 4);/* Successfully matched with
@{f(N, a)} and failed to match with @{f(out: N, a)} */
    f(out: N, 4);/* Successfully matched with
@{f(N, a)} and @{f(out: N, a)} */
}
```

• If $node1$ or $node2$ is the access path of a variable, the variable pointed to by the path is taken to replace the original node for comparison, and the comparison results meet one of the above four conditions.

Perform code replacement or function binding on $e_i$, specifically:

If $fun$ is function returning expression, mark its function declaration expression as $e_{funAnnounce}$, the expression returned by the function is $e_{funReturn}$, and the weight of the function is $w_{fun}$; Then the code segment of $e_j$ is obtained by replacing a sub code segment in the code segment of $e_i$ that matches efunannounce with the code segment corresponding to $e_{funReturn}$ after parameter replacement; If there are multiple sub code segments matching $e_{funAnnounce}$ in the $e_i$, multiple $e_j$ are generated;

The parameter replacement operation is performed for code segment, that is, the formal parameter in $e_{funAnnounce}$ is replaced with the corresponding actual parameter in $e_i$.

When $fun$ represents a simplification operation, code segments that are separated from the expression syntax tree may be generated after code replacement using function returning expression (for example, transformation $(x+1)*a+(x+1)*b \rightarrow (x+1)*(a+b)$ will make an $x+1$ separate from the syntax tree); As shown in Figure 1:

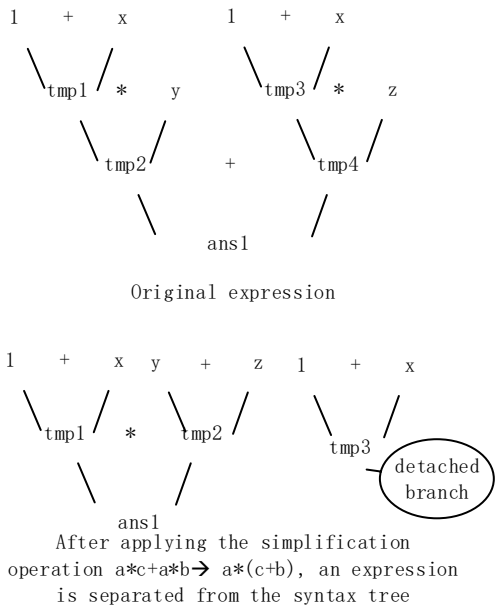

Figure 1. Example of syntax tree branch breaking away from syntax tree

When $fun$ represents a complex operation, the syntax tree of the expression may generate a ring structure after using function returning expression for code replacement (for example, transformation $(a+b)*(x+1) \rightarrow a*(x+1)+b*(x+1)$. When the syntax tree has only one $x+1$ branch, $a*(x+1)$ and $b*(x+1)$ will reference the root node of the same $x+1$ branch, and the secondary reference to $x+1$ will generate a ring structure); As shown in Figure 2:

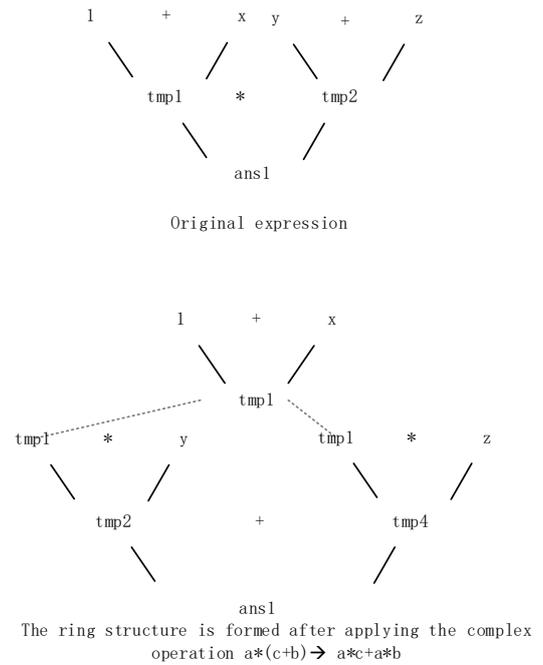

Figure 2. Example of an expression syntax tree forming a ring structure

Based on this, after the code replacement operation, you need to find and delete the code segment that is separated from the expression syntax tree in the code segment, find the ring structure and expand it; To find the ring structure, check whether the new replacement part of the syntax tree refers to the same syntax tree branch at least twice. If this happens, there is a ring structure; To expand the ring structure, copy the code segment corresponding to the syntax tree branch whose root node is referenced at least twice, replace the code address of the code segment corresponding to the copied syntax tree branch and the address used to store the intermediate variables generated at runtime with the unused address, so that the code segment copied by this syntax tree branch can be inserted into the code segment after the code replacement operation and located before the second reference position of the root node of the primitive syntax tree branch without address conflict; then insert it into the code segment after the code replacement operation, and change the second reference to the root node of the syntax tree branch causing the ring structure to the reference to the root node of the copied code segment of the inserted syntax tree branch; Repeat the operation of finding and expanding the ring structure until there is no ring structure in the new replacement part of the syntax tree.

B-7-2-4. The matching process ends when one of the following conditions is met:

• When $e_j$ satisfies that all codes have functions bound to it, the matching process ends successfully, and $e_j$ is used to replace the corresponding code segment in the code table of the longest expression where the currently executing code is located; Otherwise, add $<w_j, e_j>$ to $S(k|k+1)$ and continue the matching process:：

$$S(k|k+1) = S(k|k+1) \cup <w_j, e_j> \qquad (3)$$

• When the number of matching rounds $k$ exceeds the upper limit $kmax$ specified by the user, the matching process ends in failure;

• When there are no available elements to match in this round and the next round, it ends with failure:

$$S(k-1|k) = \emptyset \wedge S(k|k+1) = \emptyset \qquad (4)$$

Figure 3 shows the binding of expression and function of code 22 (excerpted from code 15) after pre execution:

**CODE 22** Snippet of complete code example 1
```
@get result from (x) and (y){
    new: a = x + y;
    $x + 1 == y;
    new: z = 0;
    1 * $z^2 + x * z + y == 100;
    return: a + x + z;
}
```

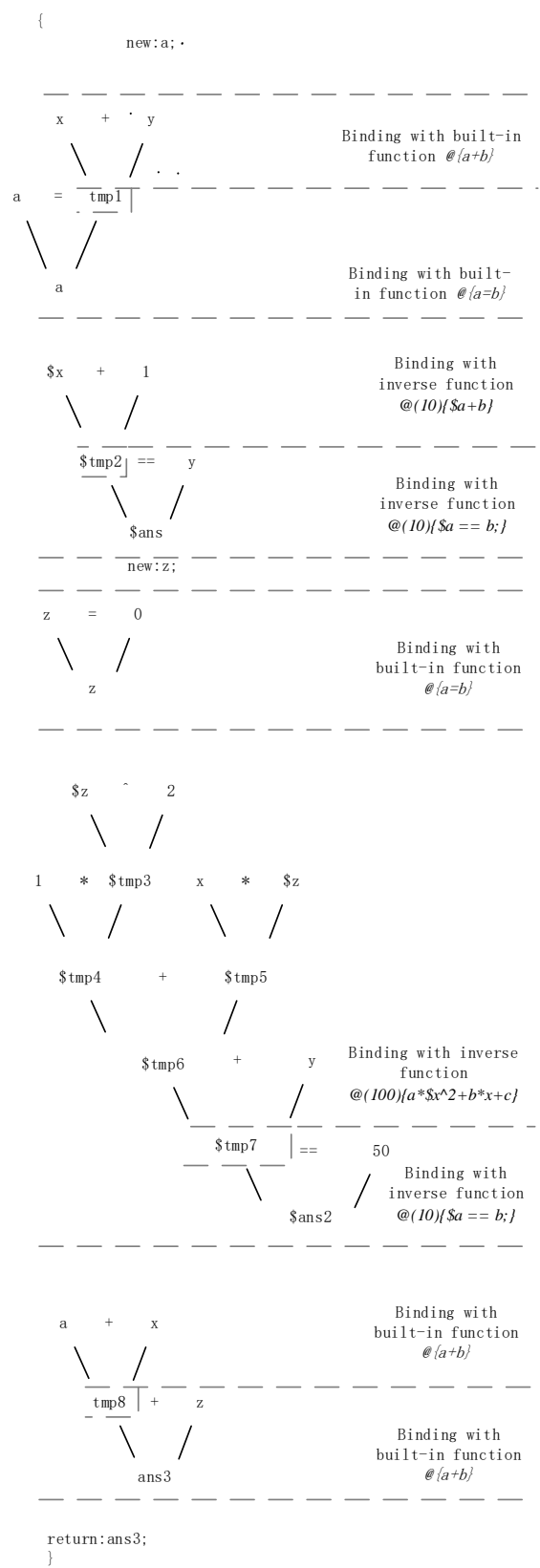

Figure 3. Binding of expressions and functions in code 22 after pre execution

Sub step B-8: when code type flag bit is "end of expression", it can be known that the longest expression of the last line of code in the currently executed code has ended, and the currently executed code is skipped;

Sub step B-9: when code type flag bit is "derivation function", use the function implementation of forward function to derive the function implementation of reverse function;

The derivation of the reverse function implementation using the forward function implementation requires the following sub steps:

Sub step b-9-1, disassemble the function body code segment of the forward function into code blocks;

Sub step b-9-2, use code blocks to derive the execution process of reverse function.

In sub step b-9-1, the code block is composed of sequence number, code segment and code block information. The sequence number is generated when the code block is created. The earlier the code block is created, the smaller its sequence number. The code block information includes: the type of code block, which is used to reflect the type of code in code segment, and can be "expression" "scope start" "scope end" "return" default; The code block depends on pending input parameters flag bit is used to indicate whether there are variables in code segment that depend on pending input parameters; A variable depends on pending input parameters, which means that determining the value of pending input parameters is a necessary condition for determining the value of this variable during the forward execution of function body code. Conversely, a variable is independent of the pending input parameter, which means that determining the value of the pending input parameter is not a necessary condition for determining the value of the parameter during the forward execution of the function body code.

Optionally, in sub-step B-9-1, the disassembly process progressively analyzes each line of code of the forward function's function body from front to back, skipping the current code if the code executable flag bit of the current analysis code is false; If the code executable flag bit of the current analysis code is "true" or "execute under certain conditions", perform the following steps according to the code type flag bit:

B-9-1-1, when the code type flag bit is "scope start": create a code block whose type is "scope start", whose code block depends on pending input parameters flag bit is "false" (independent) and code segment is the current analysis code; Add the code block from the end to the code block queue $B$;

B-9-1-2, when the code type flag bit is "scope end": create a code block whose type is "scope end", whose code block depends on pending input parameters flag bit is "false", and whose code segment is the current analysis code; Add the code block from the end to the code block queue $B$;

B-9-1-3, when the code type flag bit is "end of expression", skip the current code;

B-9-1-4 when the code type flag bit is "expression" and the function root node flag bit is "true", copy the code segment corresponding to the function call expression where the current code is located, and then copy the code segment to create the code block by performing the following operations in sequence:

B-9-1-4-1, replace variables in the code:

When a variable is modified, it will not be the original variable. In order to avoid confusion between the modified variable and the original variable, it is necessary to use a new variable to represent the modified variable to distinguish it from the original variable; It shall be handled according to the following rules:

• If a non temporary variable is to be determined, replace the variable with a new variable and add the new variable to set $A_{replace}$. At the same time, add the key value pair with the old variable replaced as the key and the new variable used to replace the old variable as the value to the variable replacement table; The variable replacement table is a one-to-one mapping table. When the key values are the same, the newly added key value pairs will overwrite the previously added key value pairs;

• If a non temporary variable is determined, query the variable replacement table for a new variable with this variable as the key. If a new variable exists, replace it with a new variable;

• If a variable is the left operand of the assignment expression, replace the variable with a new variable, and add the key value pair with the replaced variable as the key and the new variable as the value to the variable replacement table;

• A variable in a function call code segment can only be replaced with another variable at most;

The function call code segment, that is, all the codes in the code segment are bound to the same function, and only the function root node flag bit of the last line of code is "true", the syntax tree structure of the function call code segment is completely consistent with the syntax tree structure of the function declaration of the bound function, and executing the function call code segment will call the bound function;

B-9-1-4-2, determine the dependence of each variable in the code on the input parameters to be determined:

The code segment containing the variables that depend on the pending input parameters is used to backstep the pending input parameters, while the code segment of the variables that do not contain the pending input parameters serves as the known condition in the backstepping process; Specifically, the dependency of each variable in the code on the input parameter to be determined shall be determined and handled according to the following rules:

• If any child node of a variable in the syntax tree depends on the undetermined input parameter, the variable also depends on the undetermined input parameter;

• If a variable is a pending variable, and there is a variable that depends on the pending input parameter in its syntax tree, the variable also depends on the pending input parameter;

• If a variable is the left operand of the assignment expression and the right operand depends on the pending input parameter, the variable also depends on the pending input parameter;

• If a variable depends on a pending input parameter, but it was originally a determined variable in the function body of the forward function, change it to a pending variable (that is, make the parameter pending flag bit "true").

The temporary variable, that is, the variable used without declaration, a temporary variable shall be used to store the operation result of quaternion when used for the first time; Non temporary variables are variables declared by the user before use.

B-9-1-4-3, create a code block with the modified code segment:

Create a code block of type "expression"; if there are variables in the code segment that depend on pending input parameters, unbind all codes in the code segment from the function, and set the flag bit of the code block's dependent pending input parameters to true; if there is no variable that depends on the pending input parameter in the code segment, set the flag of the pending input parameter of the code block to false; add the code block from the end to the code block queue $B$;

Unbind the code from the function, that is, set the bound function address of the code to null and the function root node flag bit to false;

B-9-1-5, skip when code type flag bit is "expression" and function root node flag bit is "false";

B-9-1-6, when the code type flag bit is "return", create a code block whose type is "return" and code block depends on pending input parameters flag bit is "false"; Since the return value is known during backstepping process, assign the forward function return value $ans$ to the returned variable in code segment, and modify the code type "return" to "expression"; add the code block from the end to the code block queue $B$;

B-9-1-7. In other cases, copy the single line code segment where the current analysis code is located, and use this copy code segment to create a code block. The type of the code block is "default", and the code block depends on pending input parameters flag bit is "false". Add it to the code block queue $B$ from the end;

For example, disassemble code 22 into code blocks. After disassembly:

$$B = [1], [2], [3], [4], [5], [6], [7], [8], [9], \\ [10], [11], [12], [13], [14] \quad (5)$$

Where $[n]$ represents the code block with sequence number $n$, and the content of the code block is as shown in Figure 4 (left: code segment in the code block; right: sequence number, type, dependence on pending input parameters, parameter replacement operation information, function bound by code segment on the left):

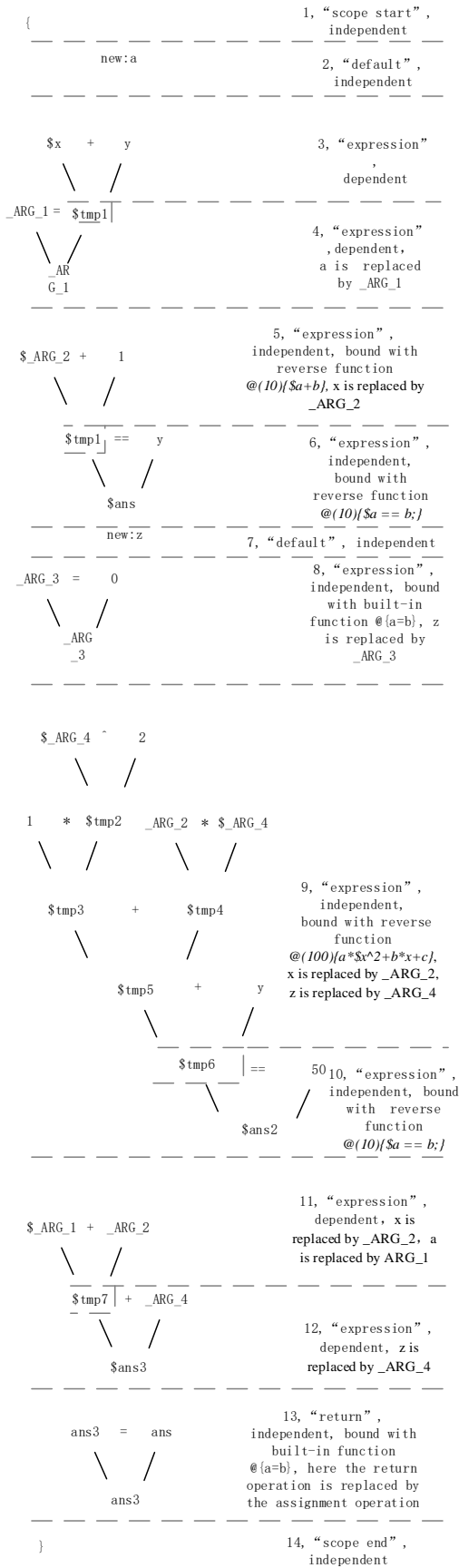

Figure 4. Contents of code blocks in *B*

In sub step B-9-2, the code block is used to deduce the execution process of reverse function. Specifically, the following operations are performed in turn:

B-9-2-1, reorganizes the code blocks in the code block queue ***B*** into the code block queue ***B**$_{depedent}$* which depends on the pending input parameters and the code block queue ***B**$_{indepedent}$* which is independent of the pending input parameters, and reorganizes the code blocks in both queues from small to large according to the sequence number of the code blocks;

B-9-2-2, when the code block queue ***B**$_{depedent}$* is not empty, the dynamic programming method is used for reverse reasoning; The following actions need to be repeated:

B-9-2-2-1, take the code block at the end of the code block queue that depends on the pending input parameters as the root node code block, backtrack and determine the dependency tree between the code blocks, and the number of nodes of the dependency tree does not exceed the maximum number determined by the user;

The child node and the parent node of the dependency tree meet the following requirements: at least one undetermined variable of the code segment in the code block of the child node is used by the code segment in the code block of its parent node; For example, the dependency tree with code block [12] as the root node and the number of nodes is 4 and its corresponding code segment are shown in Figure 5:

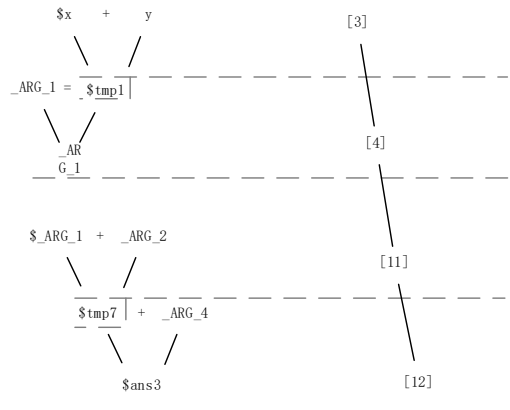

Figure 5. Dependency tree and corresponding code segment

B-9-2-2-2, select the dependency tree according to the priority, re merge each node of the selected dependency tree into a code segment, and perform code replacement and function binding operations on the code segment; If there is no dependency tree to select, the reverse inference fails;

The order of priority adopted, specifically, the less the number of necessary variables that depend on pending input parameters of the dependency tree, the higher the priority. On this basis, the less the number of nodes of the dependency tree, the higher the priority. In general, the less the necessary variables that depend on pending input parameters in the dependency tree, the more likely the code segment it is combined to bind to the function successfully. The necessary variables that depend on pending input parameters of the dependency tree, that is, variables that depend on pending input parameters and are used by the code blocks of the nodes of the dependency tree and the non dependency tree nodes in $B_{depedent}$; In contrast, if a variable that depends on the pending input parameters is only used in the nodes of the dependency tree and is not used by other code blocks in $B_{depedent}$ that do not belong to the dependency tree, then this variable is an unnecessary variable of the dependency tree that depends on the pending input parameters; For example, for the dependency tree $T$, the variable $\$\_ARG\_1$ is a necessary variable that depends on pending input parameters because it is used in code block [4], and the variables $\$tmp7$ and $\$ans3$ are non essential variables because they are only used in $T_1$.

Merge each node of the dependency tree into a code segment, that is, from the end of the leaf node of the dependency tree, merge the code segments of the code block of the leaf node into the code segments of its parent node and remove the merged successful leaf nodes from the tree until the dependency tree has only one root node, at which point the dependency tree merges successfully. In this case, the root node code segment is the code segment merged from each node of the dependency tree.

Merging process should follow the following rules:

• If the root node of the syntax tree of the code segment of the code block of the child node is an assignment expression, the syntax tree with the right value of the assignment expression as the root node (that is, the right subtree) of the code segment of the child node is used to replace all variables in the syntax tree of the code segment of the parent node that are the same as the left value of the assignment expression of the root node of the code segment syntax tree of the child node (that is, the left child node);

• If the root node of the syntax tree of the code segment of the child node is not an assignment expression, the syntax tree of the code segment of the child node is used to replace all variables in the syntax tree of the code segment of the parent node that are the same as the root node variables of the code segment syntax tree of the child node;

• When a child node is merged into a parent node, the syntax tree of the code segment of the parent node must be replaced at least once; If a replacement is not performed at all, the code block merging fails, Combining code segments using this dependency tree fails, Re execute B-9-2-2-2;

Figure 6 shows the process of merging the dependency tree $T = \{[12], [11], [4], [3]\}$ into the root node [12] (middle: dependency tree; left: code segment of the dependency tree node before merging; right: code segment of the dependency tree node after merging):

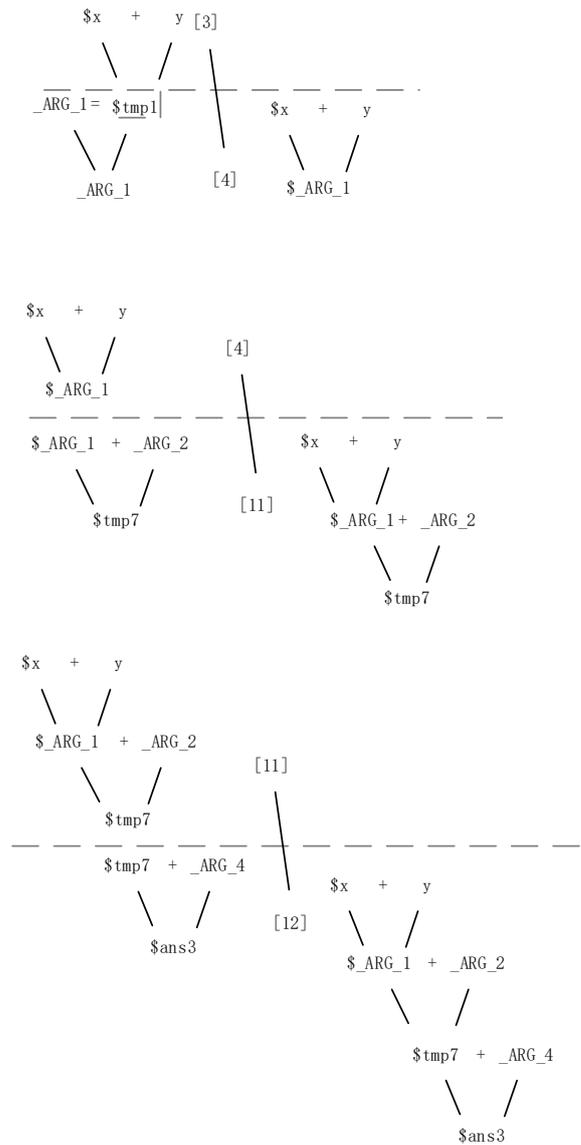

Figure 6. The process of merging the dependency tree $T$ into the root node

If a dependency tree is merged successfully, mark the original dependency tree as $T_{org}$, and the merged dependency tree is $T_{final}$, $T_{final}$ has only one node $[root]_{final}$. Perform code replacement and function binding operations on the code segment of $[root]_{final}$.

If the code replacement and function binding operations on the code segment of $[root]_{final}$ succeed, push $[root]_{final}$ from the end into the reverse execution block queue $B_{back}$; Since all pending variables in $T_{org}$ are identified after the code segments of $[root]_{final}$ are executed, all nodes in $T_{org}$, need to be removed from $B_{depedent}$ and the pending variables in other code blocks in $B_{depedent}$ need to be updated before the next round of reverse. That is, for each code block in $B_{depedent}$, if the pending variables in $T_{org}$ are used in the syntax tree of its code segment, the pending variables are changed to deterministic variables (that is, the pending flag bit of the variables is changed to false), and the flag bit of other variables in the syntax tree is modified according to the following rules:

• If all the child nodes of a variable in the syntax tree are determined, the variable is also determined;

• If a variable is the left operand of the assignment expression and the right operand is determined, the variable is also determined;

After completing the modification of the flag bit, re execute sub step B-9-2 to start the next round of backstepping;

If the code replacement operation and function binding operation fail, re execute B-9-2-2-2;

B-9-2-3, use the queue $B_{independent}$ that is independent of the code block of the pending input parameter and the queue $B_{back}$ of the reverse execution code block that is obtained by reverse inference to construct the function body code $e_{back}$ of the reverse function, specifically:

B-9-2-3-1, pop up the header code block in $B_{independent}$. The type of the popped code block is "scope start". Add the code segment of this code block to the end of the function body code segment $e_{back}$ of the reverse function; The pop-up element refers to deleting and returning a specific element from the set;

B-9-2-3-2, because new variables are used to replace the modified variables in the reverse process, which are stored in $A_{replace}$ and are not declared, you need to append code at the end of the $e_{back}$ that declares the elements in $A_{replace}$ with the code type "variable declaration";

B-9-2-3-3, if there are at least two elements in $B_{independent}$, pop up the header code block in $B_{independent}$, append the code segment of the popup code block to the end of $e_{back}$; When the type of the code block that pops up is "Expression" and there is no common temporary variable for the code segments of all elements in $e_{back}$ and $B_{independent}$, a line of code of type "End of expression" is appended to the end of the function body code segments of reverse function to indicate the end of the expression; Repeat this step until there is only one element in $B_{independent}$;

B-9-2-3-4, pop up the header code block of the reverse execution code block queue $B_{back}$, append the code table of the pop-up code block to the end of $e_{back}$; When the type of the code block that pops up is "expression" and the temporary variable used in $e_{back}$ does not exist in the reverse execution code block queue $B_{back}$, code of type "end of expression" is appended to the end of the function body code segment of reverse function to indicate the end of the expression; Repeat this step until $B_{back}$ is empty;

B-9-2-3-5, pop up the only code block in $B_{independent}$, the type of code block is "Scope end", append the code segment of this code block to the end of $e_{back}$;

B-9-2-3-6, replacing the code address of $e_{back}$ with an unused address, the address used to store temporary variables generated at runtime, and the scope address so that the body code segment $e_{back}$ of the reverse function can be inserted into the code table after the reverse function declaration position, before the next line of code, without any address conflict. After insertion, update the information of scope table and function table;

The function body, which is inversely inferred from the code block queue {B}, is structured as shown in Figure 7 (left side: function body code segment; right side: basis for generating the corresponding function body code segments on the left side, function binding operations on the left code segments, code blocks used to generate the left code segments, and so on):

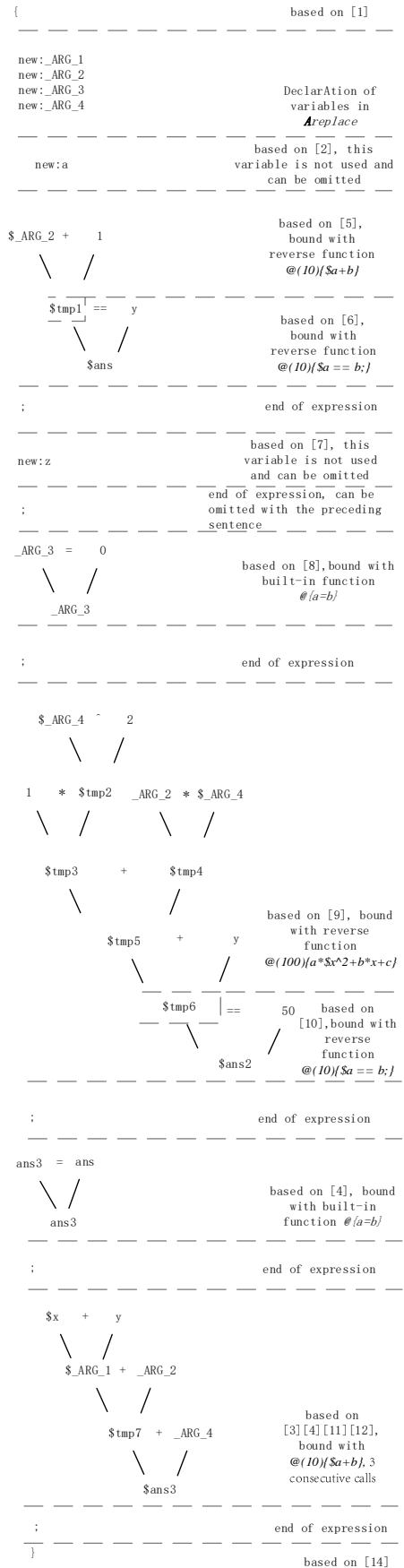

Figure 7. Function body derived from ***B***

Sub step B-10, when code type flag bit is "access member", add a key value pair with address as the key and access path as the value to the current active record data table;

Sub step B-11, when code type flag bit is "class operation", skip the current code.

The pre execution process only creates variables according to the statements to ensure that the parameter environment is correct without actual function calls and operations. At the same time, all codes in this process are only executed once at most. Compared with the simultaneous execution and reasoning, it avoids the situation that a problem may be reasoned many times.

**2.11** EXECUTE

This process will execute the extended code table generated by pre execution and get the final result. Generally, the program is still executed from front to back. However, due to the introduction of reverse function and forward function, the execution order of local code execution is not necessarily the same as the overall execution order of the program.

Specifically, the execution process includes the following steps:

Step A: skip the code whose code executable flag bit is "false";

Step B: for the code whose "code executable flag bit" is "execute under certain conditions", and this line of code is not executed by function call or condition statement jump or class instance creation, directly skip the code within the entire scope;

Step C: for the code whose "code executable flag bit" is "execute under certain conditions" and this line of code is executed by function call or condition statement jump or class instance creation, or when the code executable flag bit is "true", any one or more of the following sub steps shall be executed according to the code type:

Sub step C-1: when "code type flag bit" is "jump", if the jump condition is true, set the currently executed code to the code corresponding to the jump address and execute it; otherwise, directly skip the currently executed code;

Sub step C-2: when code type flag bit is "return operation", return the result, end the execution of this function and return to the function call location; Specifically, the returned operation value is assigned to the actual return parameter corresponding to the formal return parameter in the formal parameter and actual parameter cross reference table. If the actual return parameter does not exist, it will not be assigned; where formal return parameter refers to the corresponding parameter of the root node in function declaration expression, and actual return parameter refers to the corresponding parameter (variable) of the root node of the corresponding expression of code segment bound to this function, as shown in Figure 8:

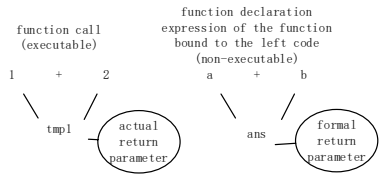

Figure 8. Examples of actual return parameter and formal return parameter

Mark the current active record as the active record to be destroyed, mark the corresponding code of return address of current active record as the currently executed code, mark the return to active record of the current active record as the current active record, execute the currently executed code, and destroy the active record to be destroyed through the recursive method; The specific process of destroying active records through the recursive method is as follows: if the parent record of a record to be destroyed is not the same active record as the current active record and is not a class instance, the parent active record of the record to be destroyed will also be marked as the active record to be destroyed, and all the active records to be destroyed will be destroyed together.

Sub step C-3: when code type flag bit is "function operation", skip the currently executed code;

Sub step C-4: when "code type flag bit" is "scope start", create an empty active record in the current active record for the scope starting from the currently executed code, initialize the active record information, and set the created active record as the current active record; In particular, if the scope is a class scope and has a parent class, create the active record of the parent class and add its pointer to the array of "pointers to parameter query active records" of the current active record before executing the next line of code in the scope;

Sub step C-5, mark the current active record as the one to be destroyed when code type flag bit is "Scope end" and the current active record is not a class instance. If return to active record does not exist, destroy the current active record and the program execution ends. Otherwise, mark the code corresponding to return address of current active record as the code currently executing. Mark the return to active record of the current active record as the current active record, use the recursive method to destroy the active record, and continue executing the program; When "code type flag bit" is "Scope end" and the current active record is a class instance, the current active record is not destroyed, only the return to active record of the current active record is marked as the current active record, and the code corresponding to the return address of the current active record is marked as the code currently executed and the program continues to execute;

Sub step C-6, when code type flag bit is "variable declaration", add a key-value pair with the parameter identifier as the key and the default value of the parameter as value to the data table of the current active record;

Sub step C-7: when the code type flag bit is "expression", for the code segment of the longest expression where the current code is located, first execute the forward functions in the longest expression in forward sequence, and then execute the reverse functions in reverse, as shown in Table 1:

Table 1. An example of the execution sequence of code segments containing forward functions and reverse function calls

| Sequence in code segment | Whether the bound function is reverse function | Actual execution sequence |
| --- | --- | --- |
| 1 | Y | 5 |
| 2 | N | 1 |
| 3 | N | 2 |
| 4 | Y | 4 |
| 5 | Y | 3 |

Perform the following two operations in sequence:

Sub step C-7-1, execute backward from the starting position of the longest expression code segment. If function root node flag bit is false, skip the currently executed code; If function root node flag bit is true and the bound function is forward function, call the bound function; If the function root node flag bit is true and the bound function is reverse function, only the variables required to call the bound function are initialized and this function is not called; Until the end of the longest expression is reached;

Sub step C-7-2: execute from the end position to the start position of the longest expression code segment (that is, code segment of the longest expression). If function root node flag bit is false or the bound function is forward function, skip the currently executed code; If the function root node flag bit is true and the bound function is reverse function, the bound function will be called and the return address of the active record created for the function call will be set to the address of the code on the previous line of the function call code segment calling this reverse function in the longest expression code segment, If the start position of function call code segment calling this reverse function coincides with the start position of the code segment of the longest expression, set the return address to the address of the next line of code after the end position of the longest expression code segment.

Sub step C-8, when "code type flag bit" is "end of expression", it indicates that the longest expression of the code on the previous line of the currently executed code has ended, skipping the currently executed code;

Sub step C-9, when code type flag bit is "derivation function", skip this line of code;

Sub step C-10, when code type flag bit is "access member", add a key value pair with address as the key and access path as the value to the current active record data table;

Sub step C-11: when code type flag bit is "class operation", skip this line of code.

## 3  SUMMARY and FORECAST

COOL pioneered the use of complex expressions as function statements, which enhanced the mathematical expression ability of logic programming languages and the applicable reasoning range. It reduces the burden of user's inverse inference problem by providing the function of reverse inference in forward execution process. It is suitable for mainstream software development processes by supporting process-oriented and object-oriented. Nevertheless, COOL still has much room for improvement:

In terms of reasoning method, the cumulative weight method is essentially a controller that adjusts the code segment with the matched function as input in each round, and the cumulative weight is used as a brief feedback on the effect of code segment adjustment to adjust the matching process. In order to achieve quicker and more accurate control through this controller, more detailed state information needs to be fed back in each matching cycle, i.e. the dimension of weight needs to be expanded, and a more reasonable weight accumulation method and filtering mechanism should be developed to improve reasoning speed. For users, increasing the dimension of the weight increases the difficulty of determining the weight and destroys the simplicity, however, machine learning can be attempted to determine the multi-dimensional weight of a function. In addition, further research is needed to expand the scope of reversible inference.

In terms of execution performance, during COOL pre-execution, traversal methods are currently used to determine functions with the same structure as expressions, which incur considerable performance overhead. To improve efficiency, we need to design an appropriate sorting and retrieval method for functions that use expressions as function declarations to avoid traversing operations.

In terms of debugging, COOL can output the contents of code table silo in the reasoning process and record the transformation process experienced by each piece of code in the code table silo, through which users can "guess" what constraints or transformation rules are missing in programming. However, this also means that COOL does not perfectly address the pain point in the debugging process of constraint logic programming: that is, to give detailed directions or suggestions for improvement when reasoning fails. This may require a lot of research.

# APPENDIX 1

Visit https://github.com/coolang2022/COOLang to get the project source code.

Visit www.dreameng.tech for more information about us.

**Author1**, HAN Ji-Peng, born in 1998, master. His main research field is Programming Language. Email:coolang2022@qq.com

**Author2**, LICHEN Zhi-Hang, born in 1999, bachelor, His main research field is Biomedical Engineering. E-mail: 13885398393@163.com

**Background**

The fifth generation programming language is one of the key research directions in the field of programming languages, which has been developing for decades. Designers originally intended computer reasoning to reduce the burden of logical thinking for users, however, the fifth-generation programming language has always been unable to be widely used compared to the fourth-generation programming language because of its obscure semantics, harsh representation of problems, limited reasoning power, and slow execution. In recent years, the development of constrained logic programming and probability logic programming has made the reasoning ability of the fifth generation programming language continuously improved, and the problems that it can handle have become more and more broad. Based on the ideas of Constraint Logic Programming and Probability Logic Programming and the grammatical style of the mainstream programming languages, this paper designs a fifth generation programming language: COOL (Constraint and Object Oriented Language), which uses forward functions, inverse functions, functions returning expressions, functions returning operation values as the basis of inference, and controls the inference process through weights, and controls the use, management and modification of rules through classes. It also improves the execution speed through the pre-execution process. We hope that COOL will be widely used and tested in actual production activities.